\documentclass[iop]{emulateapj}
\usepackage{natbib}

\usepackage[applemac]{inputenc}
\usepackage[T1]{fontenc}
\usepackage[english]{babel}
\usepackage{graphicx}

\usepackage{hyperref}
\usepackage{cleveref}

\usepackage{amsmath}

\shorttitle{On the 1/f part of the solar wind spectrum}
\shortauthors{Matteini et al.}

\begin{document}


\title{On the 1/f spectrum in the solar wind and its connection with magnetic compressibility}


\author{L. Matteini\altaffilmark{1,2}, D. Stansby\altaffilmark{2}, T.S. Horbury\altaffilmark{2}, and C.H.K. Chen\altaffilmark{3,2}}
\altaffiltext{1}{LESIA, Observatoire de Paris, Université PSL, CNRS, Sorbonne Université, Univ. Paris Diderot, Sorbonne Paris Cité, 5 place Jules Janssen, 92195 Meudon, France}
\altaffiltext{2}{Department of Physics, Imperial College London, SW7 2AZ London, UK}
\altaffiltext{3}{School of Physics and Astronomy, Queen Mary University of London, London E1 4NS, UK}


\begin{abstract}
We discuss properties of Alfvénic fluctuations with large amplitude in plasmas characterised by low magnetic field compression. We note that in such systems power laws can not develop with arbitrarily steep slopes at large scales, i.e. when $|\delta \bf{B}|$ becomes of the order of the background field $|\bf{B}|$. In such systems there is a scale $l_0$ at which the spectrum has to break due to the condition of weak compressibility.
A very good example of this dynamics is offered by solar wind fluctuations in Alfvénic fast streams, characterised by the property of constant field magnitude. 
We show here that the distribution of $\delta B=|\delta \bf{B}|$ in the fast wind displays a strong cut-off at $\delta B/|{\bf B}|\lesssim2$, as expected for fluctuations bounded on a sphere of radius $B=|{\bf B}|$.
This is also associated with a saturation of the {\it rms} of the fluctuations at large scales and introduces a specific length $l_0$ above which the amplitude of the fluctuations becomes independent on the scale $l$. Consistent with that, the power spectrum at $l>l_0$ is characterised by a -1 spectral slope, as expected for fluctuations that are scale-independent. Moreover, we show that the spectral break between the 1/f and inertial range in solar wind spectra indeed corresponds to the scale $l_0$ at which $\left<\delta B/B\right>\sim1$.
Such a simple model provides a possible alternative explanation of magnetic spectra observed in interplanetary space, also pointing out the inconsistency for a plasma to simultaneously maintain $|\bf{B}|\sim$const. at arbitrarily large scales and satisfy a Kolmogorov scaling.
\end{abstract}



\section{introduction}
The magnetic field spectrum of the fast solar wind is characterised by a double power law at intermediate and large scales, with power indices -5/3 and -1 respectively \citep{Bavassano_al_1982, Denskat_Neubauer_1982, Burlaga_Goldstein_1984}. The former corresponds to the MHD turbulence inertial range (observed typically for $10^{-3}\rm{Hz}\lesssim f\lesssim 10^{-1}\rm{Hz}$, where frequencies in the spacecraft frame correspond to Doppler-shifted spatial k-vectors in the plasma frame), while the latter, so-called 1/f range (typically for $f\lesssim10^{-3}$Hz at 1 AU), is considered the energy reservoir feeding the turbulent cascade, although the origin of this range is not well understood and still under debate \citep{Matthaeus_Goldstein_1986, Velli_al_1990, Verdini_al_2012, Chandran_2018}. Note that a spectrum with index -1 is indicative of scale-independent underlying fluctuations and a long memory in the system \citep[e.g.][]{Keshner_1982}.

In this work, we investigate the possible link of spectral properties with another well-established property observed during fast streams with large amplitude Alfvénic fluctuations, namely the almost constancy of magnetic field intensity. This property is surprising because while the total amplitude of the fluctuations $\delta B=|\delta \bf{B}|$ is of the order of the field strength $B=|{\bf B}|$, the associated variations in the magnetic field intensity $\delta |{\bf B}|$ remain small at all scales: $\delta |{\bf B}| \ll \delta B \sim B$ \citep{Belcher_Davis_1971}. Note that the nearly constant magnetic pressure has an impact also on the gas pressure, leading to equally small perturbation of the plasma density during Alfvénic fast streams. 
Geometrically, it means that the tip of the magnetic field vector is approximatively constrained on a sphere of constant radius $|{\bf B|}$ \citep[e.g.][]{Bruno_al_2004}.

We discuss here how this behaviour has a direct impact on the scaling of $\delta B$ and on the shape of its distribution (PDF).
In particular, we suggest that if the plasma is characterised by a regime of small magnetic field intensity variations, then a conflict with the expected MHD turbulence spectrum at large scales can arise. 
Indeed, large amplitude fluctuations (with $\delta B\sim B$) cannot simultaneously be organised with an arbitrary power law and maintain a low magnetic compressibility at all scales. On the contrary, if relaxing the former, the latter condition imposes a saturation of $\delta B$ for scales $l$ larger than the reference scale $l_0$ at which the average level of the fluctuations reaches the mean field amplitude $B$.
As a consequence, this saturation introduces a break in the spectrum at $l_0$ and a slope of {index~-1} (or shallower) for $l>l_0$, as for fluctuations that are independent of the scale.

This simple argument, based on the phenomenological constraint $\delta |{\bf B}| \ll \delta B$ motivated by in situ observations, leads to a straightforward justification of the spectral shape that is usually observed at large scales in space plasmas. We show in this work that the existence of a 1/f spectrum in Alfvénic fast streams is indeed associated with the presence of an observational cutoff in the distribution of the fluctuations and the saturation of their mean amplitude. Moreover, the spectral break between inertial and 1/f ranges as observed at various radial distances from the Sun always corresponds to the scale $l_0$ at which the average level of fluctuations reaches $\left<\delta B/B\right>\sim1$. 

\begin{figure}
   \includegraphics[width=8.5cm]{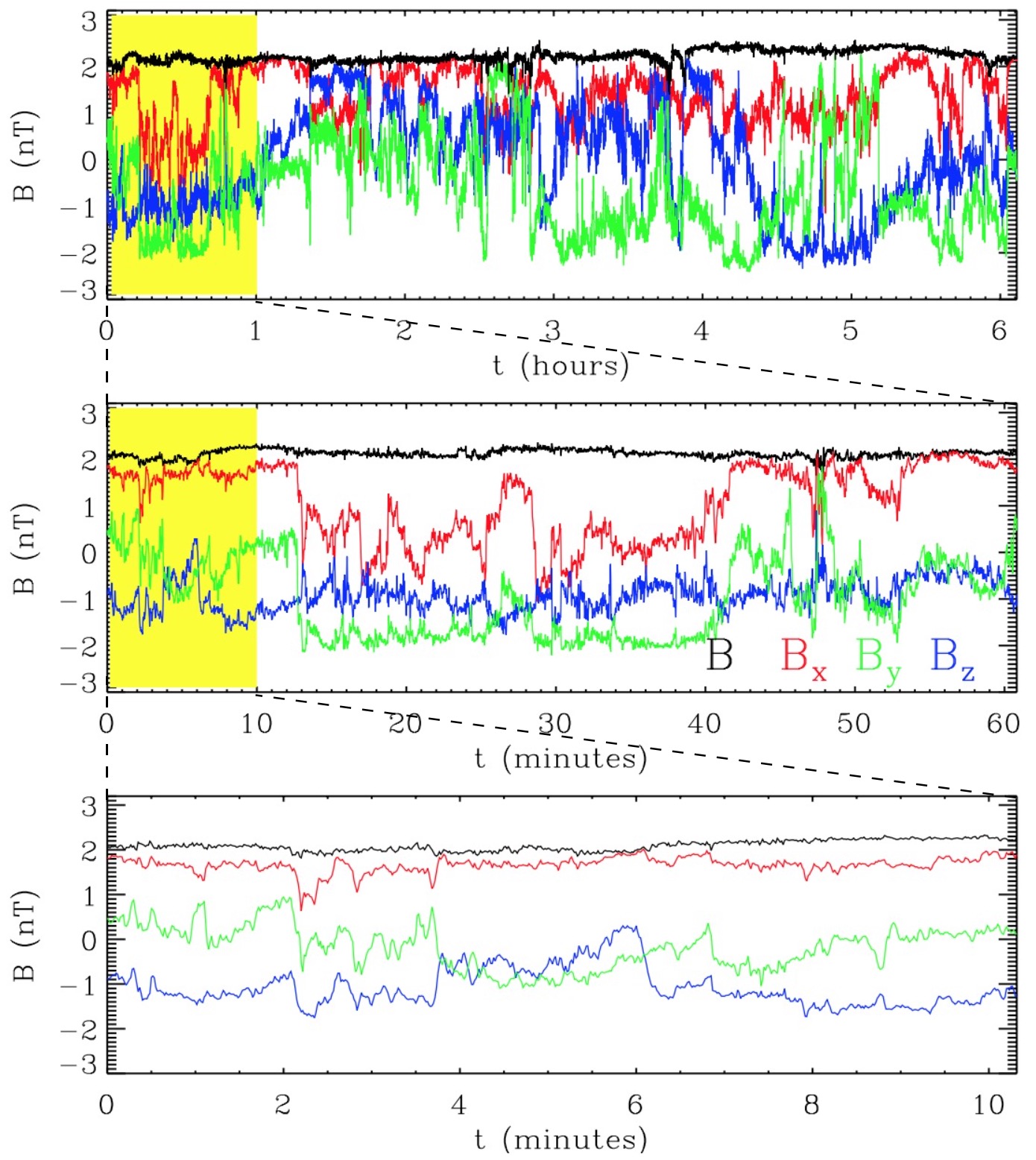} 
   \caption{Typical magnetic field fluctuations in the fast solar wind measured at various scales within the same interval (6hrs, 1hr, 10min) starting at 11:00:00 of 17/06/1995. Data are in RTN coordinate system and the magnetic field intensity is shown in black.}
   \label{fig1}\label{fig_modB}
\end{figure}

\section{Data Analysis}
We analyse Ulysses magnetic field observations over the pole during solar minimum, when the spacecraft was permanently embedded in fast wind.
We selected 150 continuous days starting from day 100 of year 1995, with time resolution of 1s, corresponding to radial distances 1.4-2.2 AU and a heliographic latitudinal variation from 30 to 80 degrees \citep{Wicks_al_2010, Chen_al_2012b}. We also use selected intervals of measurements near the ecliptic by Ulysses and by Helios at perihelion (0.3 AU).
Time increments $\delta {\bf B}(t,\Delta t)={\bf B} (t)-{\bf B}(t+\Delta t)$ are calculated using 17 logarithmically spaced lags $\Delta t$, ranging from 1 to $2\cdot10^5$s. In terms of physical scales, this covers the full MHD inertial range of the turbulence and the largest scales ($\Delta t\gtrsim10^4$s) correspond to the 1/f range \citep{Bruno_Carbone_2013}. In the following we define $\delta B$ as $|\delta {\bf B}|$ and $\delta B/B$ as $|\delta {\bf B}|/B$, where $B$ is the average intensity $|{\bf B}|$ over the interval $\Delta t$. 

\begin{figure}
   \centering
   \includegraphics[width=8.5cm]{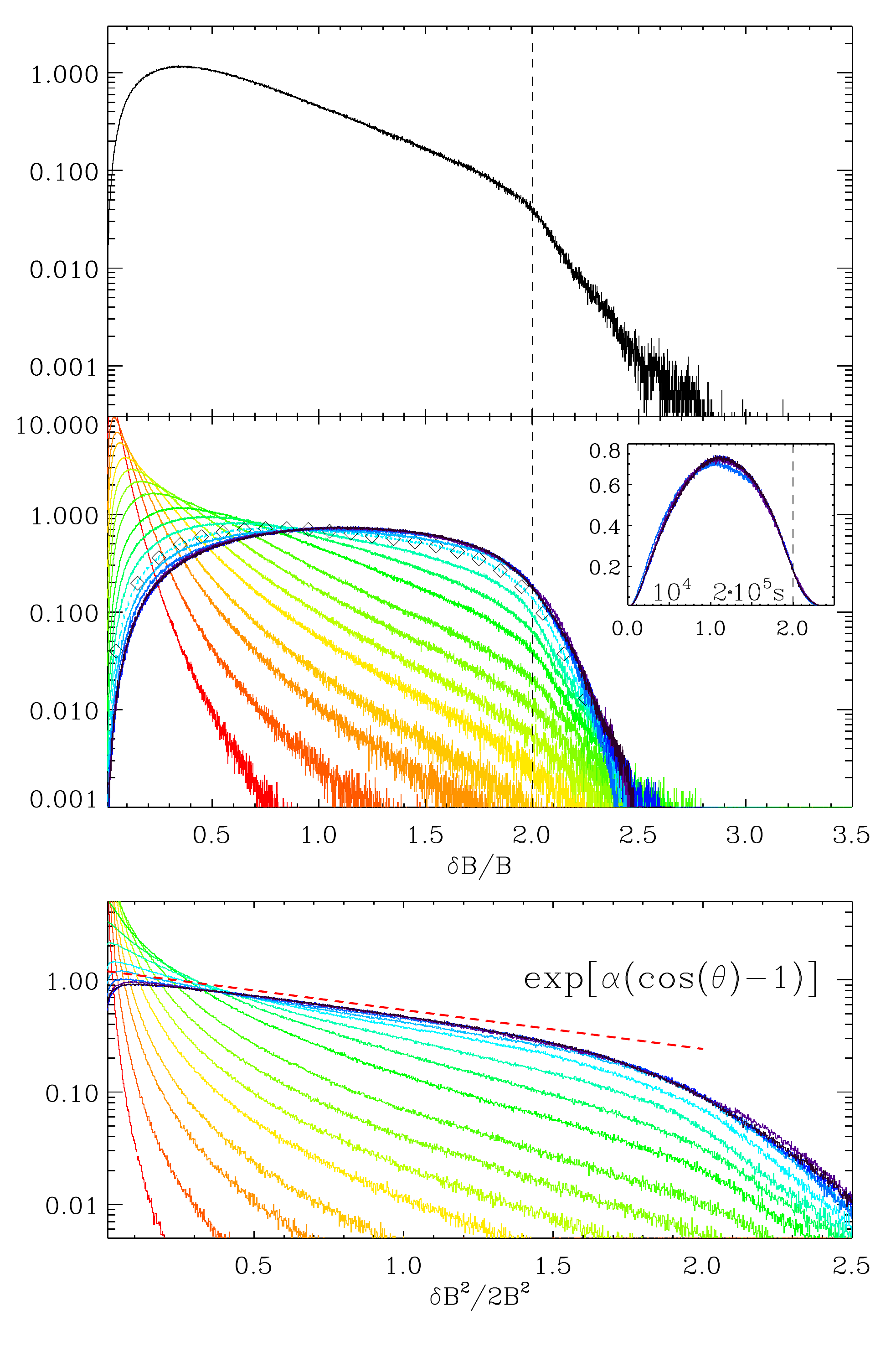} 
   \caption{Top: PDF of $\delta B/B$ at scale $\Delta t=500$s. The vertical dashed line highlights the cutoff at $\delta B/B=2$. Middle: Distributions of $\delta B/B$ over different scales, from 1s (red) to $2\cdot10^5$s (black). The PDF of $\Delta t\sim5\cdot10^3$s, roughly identifying the 1/f break scale, is highlighted by squares. The insert shows the PDFs from the 1/f range ($10^4-2\cdot10^5$s). Bottom: PDFs of $\delta B^2/2B^2$ which is directly related to the cosine of the rotation angle $\theta$. The dashed line shows the exponential dependence of Eq. (\ref{eq_angle}).}
   \label{fig_pdf_db}
\end{figure}

\section{Distribution of  $\delta B/B$}\label{observations_b}
Figure \ref{fig_modB} shows an example of typical magnetic field fluctuations in the fast solar wind taken during a period of 6hrs (top) and then within shorter intervals of 1 hour (middle) and 10 minutes (bottom). The amplitude of the fluctuations in the components decreases with the scale, going from $\delta B\sim B$ at the largest scale to $\delta B\ll B$ at smaller scales. Remarkably, $B$ (black line) is very steady and close to constant all the time ($\delta|{\bf B}|/B\lesssim10\%$), and its value is the same at all scales.

This property characterises Alfvénic fluctuations at all distances and has the geometrical consequence that the tip of the magnetic field vector moves on a sphere of approximatively constant radius \cite[e.g.][]{Bruno_al_2004, Matteini_al_2015a}.
Although this is well known \citep{Belcher_Davis_1971}, there is an important consequence of this state that has not yet been noted.
The regime of low-magnetic compressibility sets a well defined limit for the amplitude of the fluctuations:
the maximum amplitude $\delta B$ of the difference between two magnetic field measurements is twice the radius of the sphere, i.e. $\delta B \le2B$.
We can then expect that this limit is visible in the distribution of $\delta B$.

This is confirmed by the top panel of Figure~\ref{fig_pdf_db}, which shows the distribution of $\delta B/B$ at a scale of 500s, well inside the inertial range. A very clear cutoff is visible in the distribution, located at $\delta B/B\sim2$. Only fluctuations changing significantly the modulus of $\bf{B}$ (more than 10-20$\%$) populate the far righthand side of the PDF; these are mostly field depressions associated with non-Alfvénic, isolated magnetic holes \citep{Franz_al_2000}, for which $\delta B/B$ is particularly enhanced.
On the contrary the larger part of the PDF on the lefthand side ($\delta B/B\lesssim2$) constitutes the main incompressible component of the turbulence, for which $\delta |{\bf B}|\ll\delta B$.

The middle panel of Figure~\ref{fig_pdf_db} shows that the same cutoff is visible in all PDFs at large scales (green to blue), while it gradually disappears approaching kinetic scales (green to red) as expected since $\delta B\ll B$ at small scales. The PDFs at kinetic scales become narrow, corresponding to small rotations of the magnetic field vector \citep{Chen_al_2015}.

The PDF of $\Delta t\sim5\cdot10^3$s approximately corresponding to the break scale separating the inertial range from the 1/f is highlighted by black squares. Interestingly, at the largest scales (cyan to black) the distributions do not evolve further and tend to a roughly symmetric shape between $0<\delta B/B<2$, peaked at $\delta B/B \sim 1$, as highlighted in the insert panel. Moreover, at these scales the PDFs lie approximately on top of each other, as expected for a 1/f range in which the amplitude of the fluctuations becomes independent of scale. 

The last panel (bottom) shows the distribution of $\delta B^2/2B^2$ which is related to the rotation angle $\theta$ between the two magnetic field vectors. For pure rotations ${\delta B^2=2B^2-2B^2\cos{\theta}}$, so in the range [0,2] we can consider $\delta B^2/2B^2\sim 1-\cos(\theta)$ \citep{Zhdankin_al_2012}.
Moving towards large scales the distribution of $\cos(\theta)$ becomes shallower, indicating that fluctuations tend to spread over the full sphere as $\delta B\sim B$. However, even at the largest scales, corresponding to the 1/f range, the distribution does not become flat, meaning that the fluctuations do not uniformly cover the sphere and maintain a memory about the underlying mean field direction. The PDF saturates to an approximately exponential shape: 
\begin{equation}\label{eq_angle}
PDF(x)\sim e^{-\alpha x},
\end{equation}
where $x=1-\cos(\theta)$ and $\alpha$ is an empirical constant close to unity ($\alpha=0.8$). We recall that for fluctuations uniformly distributed over the sphere $\alpha=0$.

\subsection{Connection to power spectra}

\begin{figure}
   \centering
   \includegraphics[width=8.5cm]{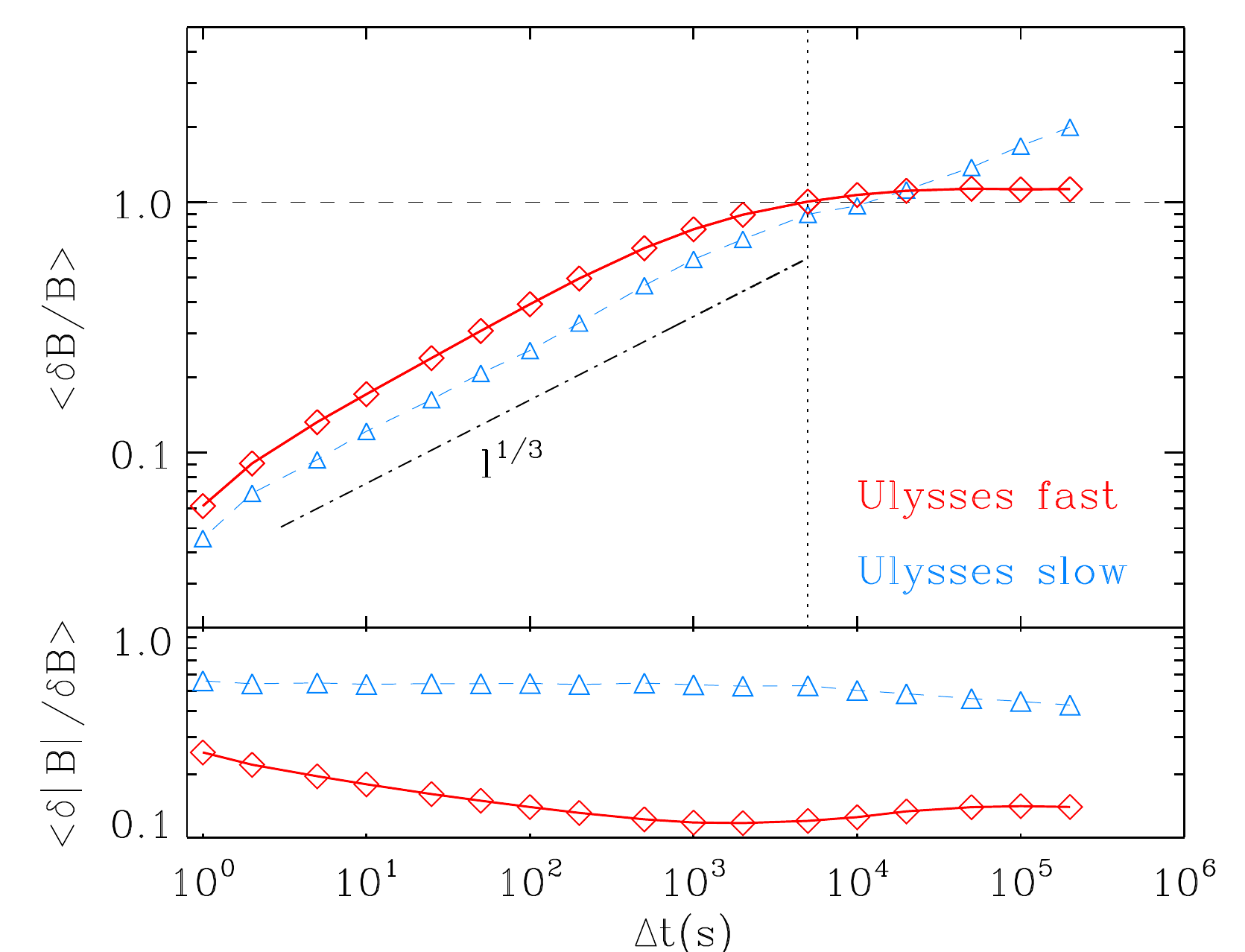} 
   \caption{Top: Average value of $\delta B/B$ over different scales, from MHD to 1/f range. In the polar wind (red diamonds) the amplitude of the fluctuations increases along the inertial range as $l^{1/3}$, consistent with a Kolmogorov scaling, and flattens as reaching the 1/f regime at $\Delta t\sim5\cdot10^3$s when $\delta B/B\sim1$. In the ecliptic slow wind (blue triangles) a 1/f range (flat distribution of $\delta B$) is not observed. Bottom: average value of magnetic compressibility $\delta |B|/\delta B$ in the two regimes.}
   \label{fig_mean_db}
\end{figure}

It is now instructive to plot the average value of $\delta B/B$ for each of the PDFs in Figure~\ref{fig_pdf_db} as a function of the scale. This is shown in the top panel of Figure~\ref{fig_mean_db} which for each scale $\Delta t$ displays the mean value of $\left<\delta B/B\right>$ (red line). This increases linearly for scales ranging from sub-minutes to $\sim1$hr, while at larger scales, $t\gtrsim2$hrs, the curve starts to flatten and saturates at $\left<\delta B/B\right>\sim1$.

As Figure~\ref{fig_modB} shows, $B$ is locally independent of the scale, and 
the distribution displayed here in practice corresponds to the distribution of the scale-dependent $\left<\delta B\right>$ normalised to the scale-independent reference $B$. 
Note that this is then analogous to simply normalising the mean value of the PDF of each scale to a constant value (the average B measured at the largest scale). As a consequence, the quantity shown in Figure~\ref{fig_mean_db} corresponds to the (normalised) first order structure function and has a direct connection to the power density spectrum of the fluctuations, thanks to the well known relation that connects $\delta B^2$ at a scale $l$ to the k-space power spectrum $P(k)$: $\delta B_l^2=P(k)\cdot k$, where $l=1/k$. In particular the slope of the power spectrum $P(k)\propto k^{-\alpha}$ is related to the exponent of the second order structure function $\delta B^2\propto l^\beta$ as: $\alpha=\beta+1$ \citep[e.g.][]{Monin_Yaglom_1975}, for $1<\alpha<3$. 

There is then a straight correspondence between the behaviour in the top panel of Figure~\ref{fig_mean_db} and the spectral slopes commonly observed in the solar wind. The increasing part with $\delta B \propto l^{1/3}$ corresponds to the inertial range of spectral index -5/3, while the flat part is the 1/f with spectral index -1. This is obviously well known; what is however new here is that the region of constant amplitude in the structure function and corresponding to the 1/f range in spectra, saturates at a normalised  value$\left<\delta B/B\right>\sim1$. This is consistent with the condition of small magnetic field compressibility discussed above. As a consequence, the spectrum breaks at the scale $l_0$ (vertical dotted line) at which the level $\left<\delta B/B\right>=1$ is reached.

\begin{figure*}[t]
   \centering
   \includegraphics[width=18cm]{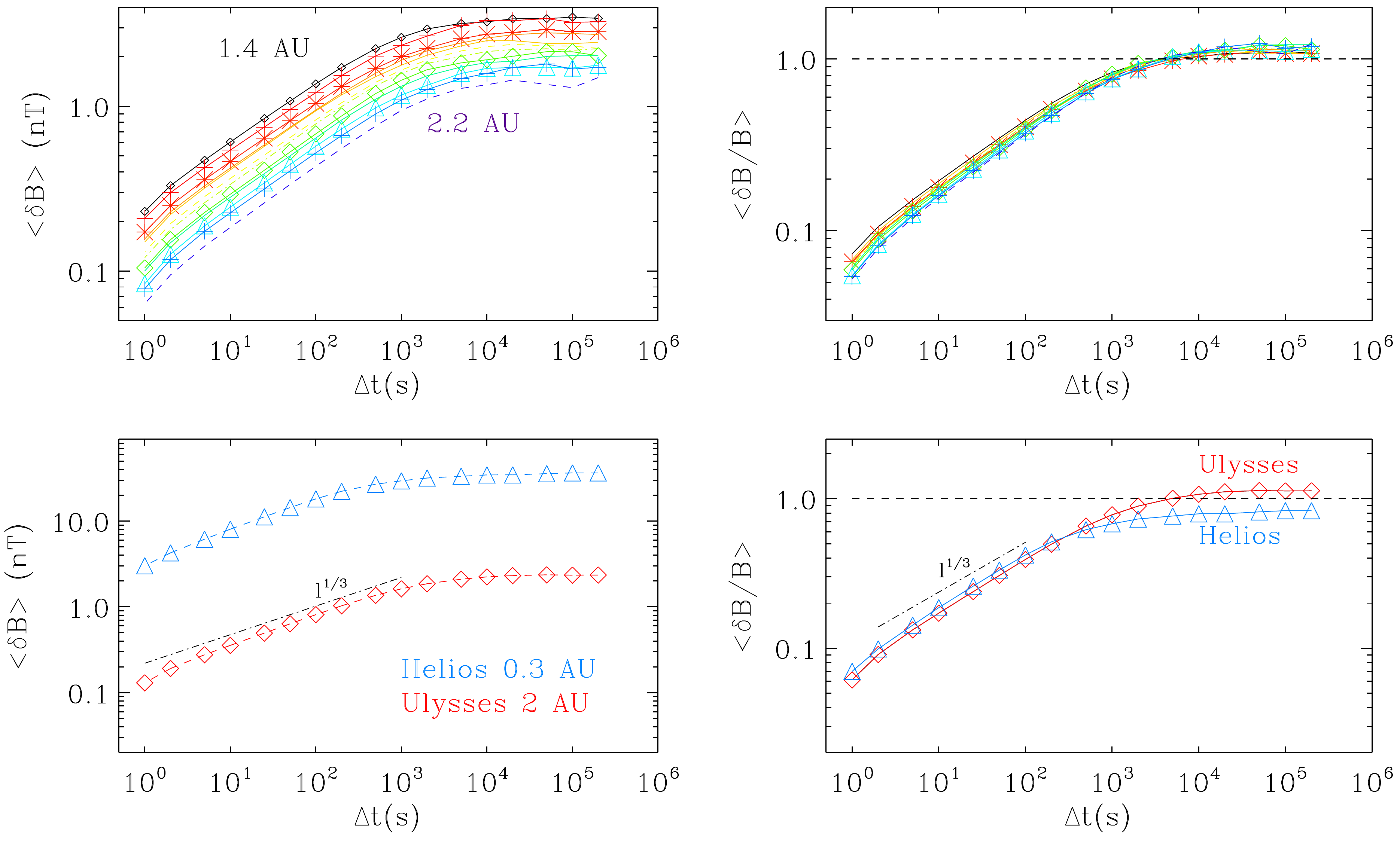} 
   \caption{Radial evolution of mean $\delta B$(nT) (left) and $\delta B/B$ (right) as a function of the scale $l$ for fast wind measurements of Ulysses over the pole (top) and compared with Helios at 0.3 AU (bottom). In the top panels different symbols/colours encode variable radial distance between 1.4 and 2.2 AU. The Kolmogorov scaling $l^{1/3}$ is shown as reference for the inertial range.}
   \label{fig_helios}
\end{figure*}

By contrast, the blue curve in the same panel shows the behaviour observed during a period of non-Alfvénic slow wind (when Ulysses was on the ecliptic, days 337-349 of year 1990). In this case the structure function does not saturate as the spectrum is not characterised by a 1/f range.
The bottom panel of the figure shows the level of magnetic compressibility, $\delta |{\bf B}|/\delta B$, in the two cases: this is small in the Alfvénic fast wind (with a minimum value $\sim0.1$ in the 1/f range), while is substantially larger in the slow wind where the condition $|\bf{B}|\sim\rm{const}$ is not observed \citep{Grappin_al_1991}.

The above picture is consistent with the idea that the constraint of low magnetic field compression sets a limiting value of $\delta B/B$.
This implies that spectra with arbitrary $\delta B/B$ at large scales cannot be realised within a condition of low magnetic field compressibility. If the latter is well satisfied, as in fast solar wind observations, then it implies that there is a scale $l_0$ such that for $l>l_0$ a steep power spectrum of fluctuations \emph{cannot} be maintained. 
In other words, this means that there is an inconsistency between extending a -5/3 inertial range to arbitrarily large scales and the fluctuations being nearly incompressible in B.
If the level of fluctuations saturates at $\left<\delta B/B\right>\sim1$ for $l>l_0$, so that the amplitude becomes independent on the scale $l$, then a spectrum with {index~-1} is the steepest possible realisation.

\section{Radial evolution}
To confirm the behaviour described above, we test the results of Figure~\ref{fig_mean_db} against radial variations. The top left panel of Figure~\ref{fig_helios} shows the scale dependent average amplitude of the fluctuations for different subintervals in the Ulysses dataset, corresponding to different radial distances. The amplitude $\delta B$ decreases with distance. The small radial variation (from 1.4 to 2.2 AU) does not introduce a major change in the break scale, although one can see that the flat region (1/f range in the spectrum) is shifted to a slightly longer time scale as known \citep[e.g.][]{Bruno_Carbone_2013}.
The right top panel shows the same curves where the amplitude is normalised to the local magnetic field strength $B$; when normalised, all spectra collapse on the same curve as in Figure~\ref{fig_mean_db}, indicating that, within the radial excursion explored by Ulysses, fluctuations populating the 1/f region are maintained at the saturation level $\delta B/B\sim1$ all along the expansion. Moreover, the break between the inertial and 1/f ranges is always identified by the scale $l_0$ at which the rms of $\delta B/B$ approaches 1.

To test further our analysis, we compare the Ulysses results at $R>1$AU with data of the Helios spacecraft, which approached the Sun as close as 0.3AU, providing thus a much larger radial excursion. We have selected a 5-day period of 1976 \citep{Bruno_al_1985}, when Helios continuously observed a highly fast stream at 0.3AU.
Data from both spacecraft are shown in the bottom panels of Figure~\ref{fig_helios} (Helios in blue, Ulysses in red). The bottom-left panel highlights the large variation in the mean amplitude values due to the large radial separation between the two measurements. In this case, we can clearly distinguish the change in the break characteristic scale, which moves to longer periods as the wind expands \citep{Bavassano_al_1982, Horbury_al_1996}. 
Also in this case, when the amplitude rms is normalised to the mean field the two curves are brought closer together, almost overlapping and both saturating approximatively at the expected value $\delta B/B\sim1$. 

\section{Discussion}
\begin{figure}[t]
   \centering
   \includegraphics[width=8.5cm]{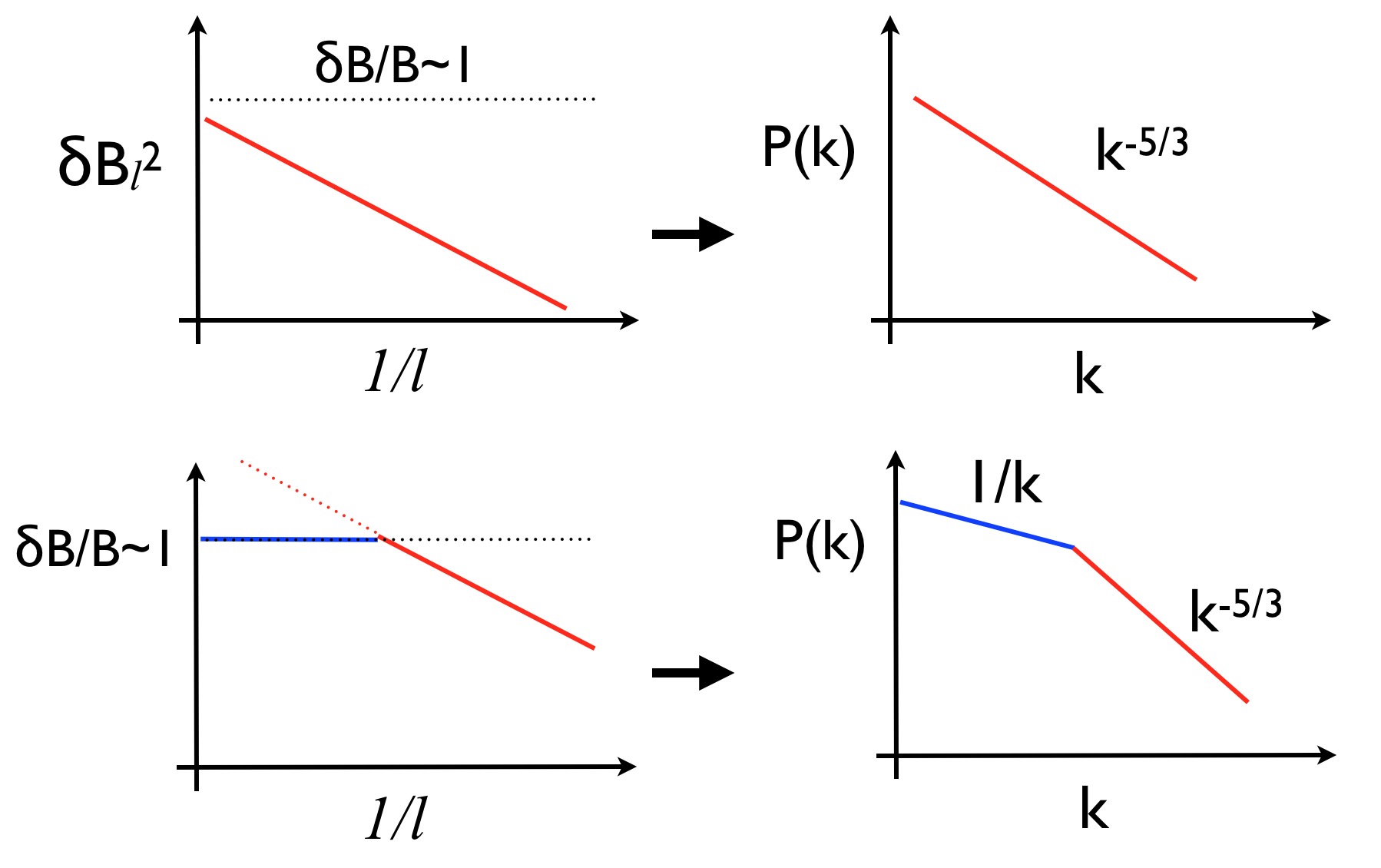} 
   \caption{Cartoon of the distribution of the fluctuations amplitude $\delta B$ over scales $l$ (left) and the corresponding power density spectra $P(k)$ (right).  A case with $\delta B\ll B$ is shown in the top panels, while a case with $\delta B/B\sim1$ is shown in the bottom. The blue sections in the bottom panels encode the part of the spectrum which is modified by the constraint of low magnetic field compression, leading to the 1/f range.}
   \label{fig_psp}
\end{figure}

A possible interpretation of the dynamics discussed here is sketched in Figure~\ref{fig_psp}. If the amplitude of the fluctuations $\delta B$ at each scale $l$ is always lower than the critical threshold $\delta B/B\sim1$ (top left panel), then a standard Kolmogorov-like spectrum $P(k)$ of slope $-5/3$ can be formed at all scales (top-right). However, if part of the fluctuations exceeds the maximum amplitude (dotted red section in the left bottom panel), the system can maintain a small magnetic field compression only somehow removing the amplitude excess and imposing a constant saturated amplitude over all the scales reaching the $\delta B/B\sim1$ condition (blue section of the curve). This situation then leads to a spectrum with two spectral slopes (bottom right), with -1 for the largest scales.

Obviously, this qualitative picture does not clarify how the plasma imposes and maintains a low magnetic compressibility, nor the physical processes that are responsible of the removal of the most compressible part of the fluctuations; these are certainly interesting and challenging questions for future theoretical studies.
However, such a simple scheme has possibly a direct application to the solar wind. Once accelerated, the absolute level of fluctuations in the solar wind is predicted to be maximum around the Alfvénic point; however, if compared to the mean field $\delta B/B \sim0.1$-$0.2$ \citep[e.g.,][]{Cranmer_VanBallegooijen_2005, Verdini_Velli_2007}. One can then expect that the turbulence can fully develop as in the top panels of Figure~\ref{fig_psp} and without relevant effects related to the low compressibility limit. As the wind expands, the relative amplitude of the fluctuations $\delta B/B$ increases and reaches unity at 0.3 AU (Helios), leading to the scenario of the bottom panels. 

Consequently, if the existence of a 1/f region is really related to the presence of a cutoff in the amplitude of the fluctuations and a saturation of their mean amplitude as suggested by Figure~\ref{fig_pdf_db}, it is then possible that this range forms during solar wind expansion somewhere in between the Alfvénic point and 0.3 AU. 
From 0.3 AU onward the plasma lies continuously on the condition $\delta B/B\sim1$ at large scales \citep[e.g.][]{Mariani_al_1978, Behannon_1978}, supporting further the idea of a saturation of the fluctuation amplitude. Outside 0.3 AU the width of the 1/f is observed to decrease with radial distance, as expected for a WKB expansion at large scales and a faster decay for the inertial range \citep[e.g.][]{Tu_al_1995}; the break scale then moves to lower frequencies, perhaps also due to some coupling between Alfvénic and compressive fluctuations \citep{Bavassano_al_2000, Malara_al_2001, DelZanna_al_2015}.
However, the evolution could be different inside 0.3 AU, before the amplitude saturation; according to our model, moving towards a lower $\delta B/B$ level closer to the Sun would also imply a reduction of the extent of the 1/f range.

\section{Conclusion}\label{conclusion}
In this work we have proposed that the 1/f region of the fast solar wind magnetic field spectrum could be generated by a saturation of the fluctuation amplitude at large scales imposed by the constraint $|\bf{B}|=\rm{const.}$, which is well verified in Alfvénic streams.
There are other models in the literature for the origin of the 1/f range \citep[e.g.][]{Matthaeus_Goldstein_1986, Velli_al_1989, Ruzmaikin_al_1996, Verdini_al_2012, Chandran_2018}. The one described in this work constitutes an alternative scenario; further testing against experimental data is needed in order to discriminate among different models.
However, intriguingly, the simple mechanism discussed above has the advantage of explaining several observational properties listed below, without the need of further assumptions.

First, and mainly, it explains the difference between the high speed Alfvénic streams where a 1/f regime is ubiquitously observed  and the typical slow wind where such a feature is absent \citep[see e.g.][]{Bruno_Carbone_2013}. The slow solar wind has commonly a lower level of power in the fluctuations, and due to its more irregular and compressible nature with respect to the fast wind, is not characterised by a $|\bf{B}|=\rm{const.}$ condition, see Figure~\ref{fig_mean_db}. 

As a further confirmation, there are periods of slow solar wind which are particularly Alfvénic \citep{Damicis_Bruno_2015}. During those periods the plasma shows properties very  similar to fast Alfvénic streams, including a higher level of fluctuations ($\delta B/B\sim1$) and a low magnetic field compression. It is then not surprising, following what has been discussed in this work, that the Alfvénic slow wind also displays a spectrum with a 1/f range, breaking at a scale $l_0$ similar to that of fast streams \citep{Damicis_al_2018}.

The 1/f part of the magnetic field spectrum has no counterpart in all fluctuating fields \citep{Tu_al_1989, Bruno_al_1996, Wicks_al_2013}. In particular, when decomposing the turbulent fluctuations using the Elsasser variables, only the dominant outward component ($\delta z^+$) shows a spectrum with 1/f range, while the inward ($\delta z^-$) does not. This can be easily explained according to the model proposed in this work. As only the outward component of the fluctuations really reaches a high enough level to satisfy $\delta B/B\sim1$, only its spectrum has a break at $l_0$. Viceversa, the lower power in the inward component keeps its spectrum below the threshold for the formation of a 1/f range and displays thus a more extended -5/3 inertial range. 

A consequence of this model is that, unless the 1/f range is formed in the Corona and just advected in interplanetary space preserving its shape, it should gradually disappear moving closer to the Sun where ${\delta B/B<1}$. It will be possible to test this prediction soon with Parker Solar Probe measurements as close as the Alfvén radius.

Finally we note that a similar saturation of Alfvénic fluctuations could be at work also in other space and astrophysical plasmas with $\delta B/B\sim1$, leading to spectra shallower than Kolmogorov at large scales \citep[e.g.][]{Hadid_al_2015}. 

\begin{acknowledgments}
Authors acknowledge valuable discussions with S. Landi, R. Bruno, A. Verdini, and M. Velli.
This work was supported by the Programme National PNST of CNRS/INSU co-funded by CNES.
TH is supported by STFC grant ST/N000692/1, DS is supported by STFC studentship ST/N504336/1,
CHKC is supported by STFC Ernest Rutherford Fellowship ST/N003748/2.
\end{acknowledgments}

\bibliographystyle{apj.bst}


\end{document}